\documentclass[]{article}
\usepackage[T1]{fontenc}
\usepackage{sectsty,xcolor}
\usepackage{amssymb,amsmath}
\usepackage{ifxetex,ifluatex}
\usepackage{fixltx2e} 
\IfFileExists{upquote.sty}{\usepackage{upquote}}{}
\ifnum 0\ifxetex 1\fi\ifluatex 1\fi=0 
  \usepackage[utf8]{inputenc}
\usepackage[left=1in,right=1in,bottom=1in,footskip=0.25in]{geometry}
\else 
  \ifxetex
    \usepackage{mathspec}
    \usepackage{xltxtra,xunicode}
  \else
    \usepackage{fontspec}
  \fi
  \defaultfontfeatures{Mapping=tex-text,Scale=MatchLowercase}
  
\fi
\IfFileExists{microtype.sty}{\usepackage{microtype}}{}
\usepackage{graphicx}
\ifxetex
  \usepackage[setpagesize=false, 
              unicode=false, 
              xetex]{hyperref}
\else
  \usepackage[unicode=true]{hyperref}
\fi
\hypersetup{breaklinks=true,
            bookmarks=true,
            pdfauthor={A. J. Landahl; D. S. Lobser; B. C. A. Morrison; K. M. Rudinger; A. E. Russo; J. W. Van Der Wall; P. Maunz},
            pdftitle={Jaqal™, the Quantum Assembly Language for QSCOUT},
            colorlinks=true,
            citecolor=blue,
            urlcolor=blue,
            linkcolor=linkblue,
            pdfborder={0 0 0}}
\definecolor{sandiablue}{RGB}{0,83,118}
\definecolor{linkblue}{RGB}{0,0,178}
\allsectionsfont{\color{sandiablue}}

\title{
\begingroup%
  \makeatletter%
  \providecommand\color[2][]{%
    \errmessage{(Inkscape) Color is used for the text in Inkscape, but the package 'color.sty' is not loaded}%
    \renewcommand\color[2][]{}%
  }%
  \providecommand\transparent[1]{%
    \errmessage{(Inkscape) Transparency is used (non-zero) for the text in Inkscape, but the package 'transparent.sty' is not loaded}%
    \renewcommand\transparent[1]{}%
  }%
  \ifx\svgwidth\undefined%
    \setlength{\unitlength}{277.19716951bp}%
    \ifx\svgscale\undefined%
      \relax%
    \else%
      \setlength{\unitlength}{\unitlength * \real{\svgscale}}%
    \fi%
  \else%
    \setlength{\unitlength}{\svgwidth}%
  \fi%
  \global\let\svgwidth\undefined%
  \global\let\svgscale\undefined%
  \makeatother%
  \begin{picture}(1,0.40461607)%
    \put(0,0){\includegraphics[width=\unitlength,page=1]{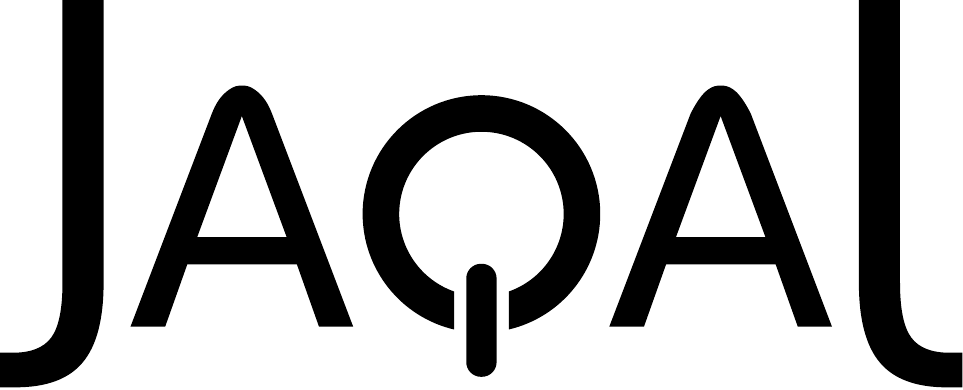}}%
    \put(3.08243825,0.18459907){\color[rgb]{0,0,0}\makebox(0,0)[lt]{\begin{minipage}{0.20086976\unitlength}\raggedright \end{minipage}}}%
  \end{picture}%
\endgroup%
~\\[1cm]\color{sandiablue} Jaqal™, the Quantum Assembly Language for QSCOUT}
\author{A. J. Landahl \and D. S. Lobser \and B. C. A. Morrison \and K. M. Rudinger \and A. E. Russo \and J. W. Van Der Wall \and P. Maunz}
\date{\vspace*{-1.5em}{\small \emph{Sandia National Laboratories, Albuquerque, NM 87185}}\vspace*{1.5em}\\March 4, 2020}

\begin{document}
\maketitle

\section{To Learn More}\label{to-learn-more}

To learn more about QSCOUT and the Jaqal™ language developed for it,
please visit \href{https://qscout.sandia.gov}{qscout.sandia.gov} or send
an e-mail to \href{mailto:qscout@sandia.gov}{qscout@sandia.gov}.

\section{Introduction}\label{introduction}

QSCOUT is the Quantum Scientific Computing Open User Testbed, a
trapped-ion quantum computer testbed realized at Sandia National
Laboratories on behalf of the Department of Energy's Office of Science
and its Advanced Scientific Computing Research (ASCR) program. As an
open user testbed, QSCOUT provides the following to its users:

\begin{itemize}
\itemsep1pt\parskip0pt\parsep0pt
\item
  \textbf{Transparency}: Full implementation specifications of the
  underlying native trapped-ion quantum gates.
\item
  \textbf{Extensibility}: Pulse definitions can be programmed to
  generate custom trapped-ion gates.
\item
  \textbf{Schedulability}: Users have full control of sequential and
  parallel execution of quantum gates.
\end{itemize}

\subsection{QSCOUT Hardware 1.0}\label{qscout-hardware-1.0}

The first version (1.0) of the QSCOUT hardware realizes a single
register of qubits stored in the hyperfine clock states of trapped
${}^{171}$Yb${}^+$ ions arranged in a one-dimensional chain. Single and multi-qubit
gates are realized by tightly focused laser beams that can address
individual ions. The native operations available on this hardware
include the following:

\begin{itemize}
\itemsep1pt\parskip0pt\parsep0pt
\item
  Global preparation and measurement of all qubits in the $z$ basis.
\item
  Parallel single-qubit rotations about any axis in the equatorial plane
  of the Bloch sphere.
\item
  The Mølmer--Sørensen two-qubit gate between any pair of qubits, in
  parallel with no other gates.
\item
  Single-qubit $Z$ gates executed virtually by adjusting the reference
  clocks of individual qubits.
\end{itemize}

Importantly, QSCOUT 1.0 does not support measurement of a subset of the
qubits. Consequently, it also does not support classical feedback. This
is because, for ions in a single chain, the resonance fluorescence
measurement process destroys the quantum states of all qubits in the ion
chain, so that there are no quantum states onto which feedback can be
applied. Future versions of the QSCOUT hardware will support feedback.

QSCOUT 1.0 uses \textbf{\emph{Just Another Quantum Assembly Language
(Jaqal)}} (described \hyperref[jaqal-quantum-assembly-language]{below})
to specify quantum programs executed on the testbed. On QSCOUT 1.0,
every quantum computation starts with preparation of the quantum state
of the entire qubit register in the $z$ basis. Then it executes a
sequence of parallel and sequential single and two-qubit gates. After
this, it executes a simultaneous measurement of all qubits in the $z$
basis, returning the result as a binary string. This sequence of
prepare-all/do-gates/measure-all can be repeated multiple times in a
Jaqal program, if desired. However, any adaptive program that uses the
results of one such sequence to issue a subsequent sequence must be done
with metaprogramming, because Jaqal does not currently support feedback.
Once the QSCOUT platform supports classical feedback, Jaqal will be
extended to support it as well.

\section{Gate Pulse File}\label{gate-pulse-file}

The laser pulses that implement built-in or custom trapped-ion gates are
defined in a \textbf{\emph{Gate Pulse File (GPF)}}. Eventually, users
will be able to write their own GPF files, but that capability will not
be available in our initial software release. However, users will be
free to specify composite gates by defining them as sub-circuit
\hyperref[macro-statement]{macros}. Additionally, custom native gates
can be added in collaboration with Sandia scientists by specifying the
pulse sequences that have to be applied to the trapped ion qubits to
realize the gate.

We have provided a GPF file for the built-in gates on the QSCOUT 1.0
platform. This file is not intended to be modified by users, so we are
not specifying its contents here. However, a full specification of the
built-in gates will be available to users of the QSCOUT 1.0 platform.
This GPF file contains pulse-level gate definitions for the QSCOUT 1.0
built-in gates listed below. All angle arguments in this list are in the
units of radians, with 40 bits of precision. The chirality of rotations
is determined using the right-hand rule.

\begin{itemize}
\itemsep1pt\parskip0pt\parsep0pt
\item
  \texttt{prepare\_all}\\\parshape 2 0.875cm \linewidth 1.1cm
  \dimexpr\linewidth-1.1cm\relax
     Prepares all qubits in the quantum register in the $|0\rangle$
  state in the $z$ basis.
\item
  \texttt{R \textless{}qubit\textgreater{} \textless{}axis angle\textgreater{} \textless{}rotation angle\textgreater{}}\\
  Counter-clockwise rotation around an axis in the equatorial plane of
  the Bloch sphere defined by
  \texttt{\textless{}axis angle\textgreater{}}, measured
  counter-clockwise from the $x$ axis, by the angle defined by
  \texttt{\textless{}rotation angle\textgreater{}}.
\item
  \texttt{Rx \textless{}qubit\textgreater{} \textless{}rotation angle\textgreater{}}\\
  Counter-clockwise rotation around the $x$ axis, by the angle defined
  by \texttt{\textless{}rotation angle\textgreater{}}.
\item
  \texttt{Ry \textless{}qubit\textgreater{} \textless{}rotation angle\textgreater{}}\\
  Counter-clockwise rotation around the $y$ axis, by the angle defined
  by \texttt{\textless{}rotation angle\textgreater{}}.
\item
  \texttt{Rz \textless{}qubit\textgreater{} \textless{}angle\textgreater{}}\\
  Counter-clockwise rotation around the $z$ axis, by the angle defined
  by \texttt{\textless{}rotation angle\textgreater{}}.
\item
  \texttt{Px \textless{}qubit\textgreater{}}\\ Counter-clockwise
  rotation around the $x$ axis, by $\pi$. (Pauli $X$ gate.)
\item
  \texttt{Py \textless{}qubit\textgreater{}}\\ Counter-clockwise
  rotation around the $y$ axis, by $\pi$. (Pauli $Y$ gate.)
\item
  \texttt{Pz \textless{}qubit\textgreater{}}\\ Counter-clockwise
  rotation around the $z$ axis, by $\pi$. (Pauli $Z$ gate.)
\item
  \texttt{Sx \textless{}qubit\textgreater{}}\\ Counter-clockwise
  rotation around the $x$ axis, by $\pi/2$. ($\sqrt{X}$ gate.)
\item
  \texttt{Sy \textless{}qubit\textgreater{}}\\ Counter-clockwise
  rotation around the $y$ axis, by $\pi/2$. ($\sqrt{Y}$ gate.)
\item
  \texttt{Sz \textless{}qubit\textgreater{}}\\ Counter-clockwise
  rotation around the $z$ axis, by $\pi/2$. ($\sqrt{Z}$ gate.)
\item
  \texttt{Sxd \textless{}qubit\textgreater{}}\\ Clockwise rotation
  around the $x$ axis, by $\pi/2$. ($\sqrt{X}^\dagger$ gate.)
\item
  \texttt{Syd \textless{}qubit\textgreater{}}\\ Clockwise rotation
  around the $y$ axis, by $\pi/2$. ($\sqrt{Y}^\dagger$ gate.)
\item
  \texttt{Szd \textless{}qubit\textgreater{}}\\ Clockwise rotation
  around the $z$ axis, by $\pi/2$. ($\sqrt{Z}^\dagger$ gate.)
\item
  \texttt{MS \textless{}qubit\textgreater{} \textless{}qubit\textgreater{} \textless{}axis angle\textgreater{} \textless{}rotation angle\textgreater{}}\\
  The general two-qubit Mølmer--Sørensen gate. (If we let $\theta$
  represent \texttt{\textless{}rotation angle\textgreater{}} and
  $\varphi$ represent \texttt{\textless{}axis angle\textgreater{}}, then
  the gate is
  \[\exp\left(-i\left(\frac{\theta}{2}\right)(\cos \varphi X + \sin \varphi Y)^{\otimes 2}\right).\]
\item
  \texttt{Sxx \textless{}qubit\textgreater{} \textless{}qubit\textgreater{}}\\
  The XX-type two-qubit Mølmer--Sørensen gate:
  \[\exp\left(-i\left(\frac{\pi}{2}\right) X\otimes X\right).\]
\item
  \texttt{measure\_all}\\ Measures all qubits of the quantum register in
  the $z$ basis. After measurement, ions will be outside the qubit
  space. Therefore, the qubits have to be prepared again before any
  other gates can be applied.
\end{itemize}

The gate pulse definitions also include idle gates with the same
duration as the single- and two-qubit gates. These have a prefix of
\texttt{I\_}. For example an idle gate of the same duration as a
\texttt{Px} can be obtained by
\texttt{I\_Px \textless{}qubit\textgreater{}}. It is important to note
that it is not necessary to explicitly insert idle on idling qubits in a
parallel block. Explicit idle gates are meant to be used for performance
testing and evaluation.

\hyperdef{}{jaqal-quantum-assembly-language}{\section{Jaqal Quantum
Assembly Language}\label{jaqal-quantum-assembly-language}}

The open nature of the QSCOUT testbed requires a flexible
\textbf{Quantum Assembly Language (QASM)} that empowers QSCOUT users to
extend the set of native gates and fully control the execution of the
quantum program on the QSCOUT testbed. Due to the proliferation of such
languages in this fledgling field, ours is named \textbf{Just Another
Quantum Assembly Language}, or \textbf{Jaqal}.

To realize our objectives, the Jaqal QASM language fulfills the
following requirements:

\begin{itemize}
\itemsep1pt\parskip0pt\parsep0pt
\item
  Jaqal fully specifies the allocation of qubits within the quantum
  register, which \emph{cannot} be altered during execution.
\item
  Jaqal requires the scheduling of sequential and parallel gate
  sequencing to be fully and explicitly specified.
\item
  Jaqal can execute any native (built-in or custom) gate specified in
  any GPF file it references.
\end{itemize}

While Jaqal is built upon a lower-level pulse definition in GPF files,
it is the lowest-level QASM programming language exposed to users in
QSCOUT. We anticipate that users will develop their own higher-level
programming languages that compile down to Jaqal. We plan to release
Jaqal-branded metaprogramming tools after user-driven innovation at this
meta-programming level settles down.

\section{Jaqal Syntax}\label{jaqal-syntax}

A Jaqal file consists of gates and metadata making those gates easier to
read and write. The gates that are run on the machine can be
deterministically computed by inspection of the source text. This
implies that there are no conditional statements at this level. This
section will describe the workings of each statement type.

Whitespace is largely unimportant except as a separator between
statements and their elements. If it is desirable to put two statements
on the same line, a `;' separator may be used. In a parallel block, the
pipe (`\textbar{}') must be used instead of the `;'. Like the semicolon,
however, the pipe is unnecessary to delimit statements on different
lines. Both Windows and Linux newline styles will be accepted.

\subsection{Identifiers}\label{identifiers}

Gate names and qubit names have the same character restrictions. Similar
to most programming languages, they may contain, but not start with,
numerals. They are case sensitive and may contain any non-accented Latin
character plus the underscore. Identifiers cannot be any of the keywords
of the language.

\subsection{Comments}\label{comments}

C/C++ style comments are allowed and treated as whitespace. A comment
starting with `//' runs to the end of the current line, while a comment
with `/*' runs until a `*/' is encountered. These comments do not nest,
which is the same behavior as C/C++.

\subsection{Header Statements}\label{header-statements}

A properly formatted Jaqal file comprises a header and body section. All
header statements must precede all body statements. The order of header
statements is otherwise arbitrary except that all objects must be
defined before their first use.

\subsubsection{Register Statement}\label{register-statement}

A register statement serves to declare the user's intention to use a
certain number of qubits, referred to in the file with a given name. If
the machine cannot supply this number of qubits then the entire program
is rejected immediately.

The following line declares a register named \texttt{q} which holds 7
qubits.

\begin{verbatim}
register q[7]
\end{verbatim}

\subsubsection{Map Statement}\label{map-statement}

While it is sufficient to refer to qubits by their offset in a single
register, it is more convenient to assign names to individual qubits.
The map statement effectively provides an alias to a qubit or array of
qubits under a different name. The following lines declare the single
qubit \texttt{q{[}0{]}} to have the name \texttt{ancilla} and the array
\texttt{qubits} to be an alias for \texttt{q}. Array indices start with
0.

\begin{verbatim}
register q[3]
map ancilla q[0]
map qubits q
\end{verbatim}

The map statement will also support Python-style slicing. In this case,
the map statement always declares an array alias. In the following line
we relabel every other qubit to be an ancilla qubit, starting with index
1.

\begin{verbatim}
register q[7]
map ancilla q[1:7:2]
\end{verbatim}

After this instruction, \texttt{ancilla{[}0{]}} corresponds to
\texttt{q{[}1{]}}; \texttt{ancilla{[}1{]}} and \texttt{ancilla{[}2{]}}
correspond to \texttt{q{[}3{]}}and \texttt{q{[}5{]}}, respectively.

\subsubsection{Let Statement}\label{let-statement}

We allow identifiers to replace integers or floating point numbers for
convenience. There are no restrictions on capitalization. An integer
defined in this way may be used in any context where an integer literal
is valid and a floating point may similarly be used in any context where
a floating point literal is valid. Note that the values are constant,
once defined.

Example:

\begin{verbatim}
let total_count 4
let rotations 1.5
\end{verbatim}

\subsection{Body Statements}\label{body-statements}

\subsubsection{Gate Statement}\label{gate-statement}

Gates are listed, one per statement, meaning it is terminated either by
a newline or a separator. The first element of the statement is the gate
name followed by the gate's arguments which are whitespace-separated
numbers or qubits. Elements of quantum registers, mapped aliases, and
local variables (see section on \hyperref[macro-statement]{macros}) may
be freely interchanged as qubit arguments to each gate. The names of the
gates are fixed but determined in the Gate Pulse File, except for
macros. The number of arguments (``arity'') must match the expected
number. The following is an example of what a 2-qubit gate may look
like.

\begin{verbatim}
register q[3]
map ancilla q[1]
Sxx q[0] ancilla
\end{verbatim}

The invocation of a macro is treated as completely equivalent to a gate
statement.

\subsubsection{Gate Block}\label{gate-block}

Multiple gates and/or macro invocations may be combined into a single
block. This is similar, but not completely identical, to how C or
related languages handle statement blocks. Macro definitions and header
statements are not allowed in gate blocks. Additionally, statements such
as macro definitions or loops expect a gate block syntactically and are
not satisfied with a single gate, unlike C.

Two different gate blocks exist: sequential and parallel. Sequential
gate blocks use the standard C-style `\{\}' brackets while parallel
blocks use angled `\textless{}\textgreater{}' brackets, similar to C++
templates. This choice was made to not conflict with `{[}{]}' brackets,
which are used in arrays, and to reserve `()' for possible future use.
In a sequential block, each statement, macro, or gate block waits for
the previous to finish before executing. In a parallel gate block, all
operations are executed at the same time. It is an error to request
parallel operations that the machine is incapable of performing, however
it is not syntactically possible to forbid these as they are determined
by hardware constraints which may change with time.

\hyperref[loop-statement]{Looping statements} are allowed inside
sequential blocks, but not inside parallel blocks. Blocks may be
arbitrarily nested so long as the hardware can support the resulting
sequence of operations. Blocks may not be nested directly within other
blocks of the same type.

The following statement declares a parallel block with two gates.

\begin{verbatim}
< Sx q[0] | Sy q[1] >
\end{verbatim}

This does the same but on different lines.

\begin{verbatim}
<
    Sx q[0]
    Sy q[1]
>
\end{verbatim}

Here is a parallel block nested inside a sequential one.

\begin{verbatim}
{
    Sxx q[0] q[1]
    < Sx q[0] | Sy q[1] >
}
\end{verbatim}

And sequential blocks may be nested inside parallel blocks.

\begin{verbatim}
<
    Sx q[0]
    { Sx q[1] ; Sy q[1] }
>
\end{verbatim}

\subsubsection{Timing within a parallel
block}\label{timing-within-a-parallel-block}

If two gates are in a parallel block but have different durations
(\emph{e.g.}, two single-qubit gates of different length), the default
behavior is to \emph{start} each gate within the parallel block
simultaneously. The shorter gate(s) will then be padded with idles until
the end of the gate block. For example, the command

\begin{verbatim}
<
    Rx q[1] 0.1
    Sx q[2]
>
\end{verbatim}

results in the \texttt{Rx} gate on \texttt{q{[}1{]}} with angle 0.1
radians and \texttt{Sx} gate on \texttt{q{[}2{]}} both starting at the
same time; the \texttt{Rx} gate will finish first and \texttt{q{[}1{]}}
will idle while the \texttt{Sx} gate finishes. Once the Jaqal gate set
becomes user-extensible, users may define their own scheduling within
parallel blocks (\emph{e.g.}, so that gates all \emph{finish} at the
same time instead).

\hyperdef{}{macro-statement}{\subsubsection{Macro
Statement}\label{macro-statement}}

A macro can be used to treat a sequence of gates as a single gate. Gates
inside a macro can access the same qubit registers and mapped aliases at
the global level as all other gates, and additionally have zero or more
arguments which are visible. Arguments allow the same macro to be
applied on different combinations of physical qubits, much like a
function in a classical programming language.

A macro may use other macros that have already been declared. A macro
declaration is complete at the \emph{end} of its code block. This
implies that recursion is impossible. It also implies that macros can
only reference other macros created earlier in the file. Due to the lack
of conditional statements, recursion always creates an infinite loop and
is therefore never desirable.

A macro is declared using the \texttt{macro} keyword, followed by the
name of the macro, zero or more arguments, and a code block. Unlike C, a
macro must use a code block, even if it only has a single statement.

The following example declares a macro.

\begin{verbatim}
macro foo a b {
    Sx a
    Sxx a q[0]
    Sxx b q[0]
}
\end{verbatim}

To simplify parsing, a line break is not allowed before the initial
`\{', unlike C. However, statements may be placed on the same line
following the `\{'.

\hyperdef{}{loop-statement}{\subsubsection{Loop
Statement}\label{loop-statement}}

A gate block may be executed for a fixed number of repetitions using the
loop statement. The loop statement is intentionally restricted to
running for a fixed number of iterations. This ensures it is easy to
deterministically evaluate the runtime of a program. Consequently, it is
impossible to write a program which will not terminate.

The following loop executes a sequence of statements seven times.

\begin{verbatim}
loop 7 {
    Sx q[0]
    Sz q[1]
    Sxx q[0] q[1]
}
\end{verbatim}

The same rules apply as in macro definitions: `\{' must appear on the
same line as \texttt{loop}, but other statements may follow on the same
line.

Loops may appear in sequential gate blocks, but not in parallel gate
blocks.

\section{Extensibility}\label{extensibility}

As Jaqal and the QSCOUT project more broadly have extensibility as
stated goals, it is important to clarify what is meant by this term.
Primarily, Jaqal offers extensibility in the gates that can be
performed. This will occur through the gate pulse file and the use of
macros to define composite gates that can be used in all contexts a
native gate can. Jaqal will be incrementally improved as new hardware
capabilities come online and real world use identifies areas for
enhancement. The language itself, however, is not intended to have many
forms of user-created extensibility as a software developer might
envision the term. Features we do not intend to support include, but are
not limited to, pragma statements, user-defined syntax, and a foreign
function interface (i.e.~using custom C or Verilog code in a Jaqal
file).

\section{Examples}\label{examples}

\subsection{Bell state preparation}\label{bell-state-preparation}

This example prepares a Bell state using the classic Hadamard and
controlled X circuit, then measures it in the computational basis. Up to
the limits of gate fidelity, the measurements of the two qubits should
always match.

\begin{verbatim}
macro hadamard target { // A Hadamard gate can be implemented as
    Sy target           // a pi/2 rotation around Y
    Px target           // followed by a pi rotation around X.
}

macro cnot control target {  // CNOT implementation from Maslov (2017)
    Sy control               //
    Sxx control target
    <Sxd control | Sxd target>  // we can perform these in parallel
    Syd control
}

register q[2]

prepare_all         // Prepare each qubit in the computational basis.
hadamard q[0]
cnot q[1] q[0]
measure_all         // Measure each qubit and read out the results.
\end{verbatim}

However, there's a more efficient way of preparing a Bell state that
takes full advantage of the native Mølmer-Sørensen interaction of the
architecture, rather than using it to replicate a controlled-X gate. The
following snippet of code repeats that interaction 1024 times, measuring
and resetting the ions after each time. All 1024 measurement results
will be reported to the user.

\begin{verbatim}
register q[2]

loop 1024 {
    prepare_all
    Sxx q[0] q[1]
    measure_all
}
\end{verbatim}

\subsection{Single-Qubit Gate Set
Tomography}\label{single-qubit-gate-set-tomography}

\begin{verbatim}
register q[1]

// Fiducials
macro F0 qubit { I_Sx qubit }
macro F1 qubit { Sx qubit }
macro F2 qubit { Sy qubit }
macro F3 qubit { Sx qubit; Sy qubit}
macro F4 qubit { Sx qubit; Sx qubit; Sx qubit }
macro F5 qubit { Sy qubit; Sy qubit; Sy qubit }

// Germs
macro G0 qubit { Sx qubit }
macro G1 qubit { Sy qubit }
macro G2 qubit { I_Sx qubit }
macro G3 qubit { Sx qubit; Sy qubit }
macro G4 qubit { Sx qubit; Sy qubit; I_Sx qubit }
macro G5 qubit { Sx qubit; I_Sx qubit; Sy qubit }
macro G6 qubit { Sx qubit; I_Sx qubit; I_Sx qubit }
macro G7 qubit { Sy qubit; I_Sx qubit; I_Sx qubit }
macro G8 qubit { Sx qubit; Sx qubit; I_Sx qubit; Sy qubit }
macro G9 qubit { Sx qubit; Sy qubit; Sy qubit; I_Sx qubit }
macro G10 qubit { Sx qubit; Sx qubit; Sy qubit; Sx qubit; Sy qubit; Sy qubit }

// Length 1
prepare_all
F0 q[0]
measure_all

prepare_all
F1 q[0]
measure_all

prepare_all
F2 q[0]
measure_all

prepare_all
F3 q[0]
measure_all

prepare_all
F4 q[0]
measure_all

prepare_all
F5 q[0]
measure_all

prepare_all
F1 q[0]; F1 q[0]
measure_all

prepare_all
F1 q[0]; F2 q[0]
measure_all

// and many more
// Repeated germs can be realized with the loop

prepare_all
F1 q[0]
loop 8 { G1 q[0] }
F1 q[0]
measure_all
\end{verbatim}

\hyperdef{}{data-output-format}{\section{Data Output
Format}\label{data-output-format}}

When successfully executed, a single Jaqal file will generate a single
ASCII text file (Linux line endings) in the following way:

\begin{enumerate}
\def\labelenumi{\arabic{enumi}.}
\item
  Each call of \texttt{measure\_all} at runtime will add a new line of
  data to the output file. (If \texttt{measure\_all} occurs within a
  \texttt{loop} (or nested loops), then multiple lines of data will be
  written to the output file, one for each call of \texttt{measure\_all}
  during execution.)
\item
  Each line of data written to file will be a single bitstring, equal in
  length to the positive integer passed to \texttt{register} at the
  start of the program.
\item
  Each bitstring will be written in least-significant bit order (little
  endian).
\end{enumerate}

For example, consider the program:

\begin{verbatim}
register q[2]

loop 2 {
    prepare_all
    Px q[0]
    measure_all
}

loop 2 {
    prepare_all
    Px q[1]
    measure_all
}
\end{verbatim}

Assuming perfect execution, the output file would read as:

\begin{verbatim}
10
10
01
01
\end{verbatim}

While this output format will be ``human-readable'', it may nevertheless
be unwieldy to work with directly. Therefore, a Python-based parser will
be written to aid users in manipulating output data.

\section{Possible Future
Capabilities}\label{possible-future-capabilities}

Jaqal is still under development, and will gain new features as the
QSCOUT hardware advances. While the precise feature set of future
versions of Jaqal is still undetermined, we discuss some features that
may be added, and in some cases identify workarounds for the current
lack of those features.

\subsection{Subset Measurement}\label{subset-measurement}

Currently, the measurement operation of the QSCOUT hardware acts on all
ions in the trap, destroying their quantum state and taking them out of
the computational subspace. Future versions of the QSCOUT hardware will
allow for the isolation and measurement of a subset of qubits with a
command of the form
\texttt{measure\_subset \textless{}qubit\textgreater{} ...}. Similarly,
a \texttt{prepare\_subset \textless{}qubit\textgreater{} ...} operation
will allow the reuse of measured qubits without destroying the quantum
state of the remainder. These would be implemented in a Gate Pulse File,
and not require a change to the Jaqal language.

\subsection{Measurement Feedback}\label{measurement-feedback}

The QSCOUT hardware does not currently support using measurement
outcomes to conditionally execute future gates. We expect this
capability will be added in a future version of the QSCOUT hardware, and
Jaqal programs will be able to use that capability once it exists. We
have chosen to delay adding the syntax for measurement feedback to Jaqal
until that time, in order to allow us the flexibility to choose a syntax
that best allows users to take advantage of the actual capabilities of
our hardware, once those are known.

\subsection{Classical Computation}\label{classical-computation}

Jaqal does not currently support any form of classical computation. We
understand that this is a limitation, and expect future versions of
Jaqal to do so. There are two relevant forms of classical computation
that we are considering for Jaqal.

\subsubsection{Compile-Time Classical
Computation}\label{compile-time-classical-computation}

Performing classical computations at compile-time, before the program is
sent to the quantum computer, can vastly increase the expressiveness of
the language. For example, consider the following experiment,
\emph{which is not currently legal Jaqal code:}

\begin{verbatim}
register q[1]

let pi 3.1415926536

loop 100 {
    prepare_all
    Ry q[0] pi/32
    measure_all
    prepare_all
    Ry q[0] pi/16
    measure_all
    prepare_all
    Ry q[0] 3*pi/32
    measure_all
    prepare_all
    Ry q[0] pi/8
    measure_all
}
\end{verbatim}

Currently, Jaqal does not support inline parameter calculations like the
above. The recommended workaround is to define additional constants as
needed:

\begin{verbatim}
register q[1]

let pi_32   0.09817477042
let pi_16   0.1963495408
let pi_3_32 0.2945243113
let pi_8    0.3926990817

loop 100 {
    prepare_all
    Ry q[0] pi_32
    measure_all
    prepare_all
    Ry q[0] pi_16
    measure_all
    prepare_all
    Ry q[0] pi_3_32
    measure_all
    prepare_all
    Ry q[0] pi_8
    measure_all
}
\end{verbatim}

Another example of a case where compile-time classical computation could
be useful is in macro definition. For example, if you wished to define a
macro for a controlled z rotation in terms of a (previously-defined)
CNOT macro:

\begin{verbatim}
...
macro CNOT control target { ... }

macro CRz control target angle {
    Rz target angle/2
    CNOT control target
    Rz target -angle/2
    CNOT control target
}
...
\end{verbatim}

Again, the above example \emph{is not currently legal Jaqal.} We
recommend, in such cases, that you manually unroll macros as needed,
then define additional constants as above. That is, rather than using
the above macro:

\begin{verbatim}
...
let phi 0.7853981634;
...
CRz q[0] q[1] phi;
...
\end{verbatim}

You should instead call the gates the macro is made up of, substituting
the results of the appropriate calculations yourself:

\begin{verbatim}
...
let phi 0.7853981634;
let phi_2 0.3926990817;
let phi_m_2 -0.3926990817;
...
Rz q[1] phi_2; CNOT q[0] q[1]; Rz q[1] phi_m_2; RNOT q[0] q[1];
...
\end{verbatim}

We recognize that this ``manual compilation'' is a significant
inconvenience for writing readable and expressive code in Jaqal. We
expect to include compile-time classical computation in a relatively
early update to Jaqal, likely even before measurement feedback is
available. Fortunately, metaprogramming (automated code generation)
significantly eases the burden of the lack of classical computation
features, and we highly recommend it to users of Jaqal.

\subsubsection{Run-Time Classical
Computation}\label{run-time-classical-computation}

Users may also wish to do classical computation while a Jaqal program is
running, based on the results of measurements. For example, in hybrid
variational algorithms, a classical optimizer may use measurement
results from one circuit to choose rotation angles used in the next
circuit. In error-correction experiments, a decoder may need to compute
which gates are necessary to restore a state based on the results of
stabilizer measurements. Adaptive tomography protocols may need to
perform statistical analyses on measurement results to determine which
measurements will give the most information. As can be seen from the
above examples, run-time classical computation is useful only when
measurement feedback is possible. Accordingly, we will consider this
feature after we have added support for measurement feedback. However,
use cases like adaptive tomography and variational algorithms can be
implemented via metaprogramming techniques. After running a Jaqal file
on the QSCOUT hardware, a metaprogram can parse the
\hyperref[data-output-format]{measurement results}, then use that
information to generate a new Jaqal file to run.

\subsection{Randomness}\label{randomness}

Executing quantum programs with gates chosen via classical randomness is
desirable for a variety of reasons. Applications of randomized quantum
programs include hardware benchmarking, error mitigation, and some
quantum simulation algorithms. Jaqal does not currently have built-in
support for randomization, although it may in the future, likely in
combination with support for run-time classical computation. Our
currently recommended workaround is to pre-compute any randomized
elements of the algorithm, automatically generating Jaqal code to
execute the random circuit selected. For example, the following program
isn't currently possible, as there's no means of generating a random
angle in Jaqal directly:

\begin{verbatim}
register q[1]

loop 100 {
    prepare_all
    // Do an X rotation on q[0] by a random angle between 0 and 2*pi.
    measure_all
}
\end{verbatim}

However, the same effect can be obtained by a metaprogram (written in
Python, for the sake of example) that generates a Jaqal program:

\begin{verbatim}
from random import uniform
from math import pi
with open("randomness_example.jql", "w") as f:
    f.write("register q[1]\n\n")
    for idx in range(100):
        angle = uniform(0.0, 2.0 * pi)
        f.write("prepare_all\n")
        f.write("Rx q[0] %f\n" % angle)
        f.write("measure_all\n\n")
\end{verbatim}

While the generated Jaqal program is much larger than one that could be
written in a potential future version of Jaqal that supported randomized
execution, the metaprogram that generates it is quite compact.

\section{Acknowledgements}\label{acknowledgements}

This material was funded by the U.S. Department of Energy, Office
of Science, Office of Advanced Scientific Computing Research Quantum Testbed
Program.

Sandia National Laboratories is a multi-mission laboratory managed and
operated by National Technology and Engineering Solutions of Sandia,
LLC, a wholly owned subsidiary of Honeywell International, Inc., for
DOE's National Nuclear Security Administration under contract
DE-NA0003525.

\end{document}